\documentclass[nopreprintline]{elsarticle} 

\usepackage[utf8]{inputenc}
\usepackage[margin=1in]{geometry} 
\usepackage{graphicx}

\usepackage{amsmath}
\usepackage{amsthm}
\usepackage{amssymb}

\usepackage{dcolumn} 
\usepackage{textcomp} 
\usepackage{xcolor}
\usepackage[normalem]{ulem} 

\begin{document}

\begin{frontmatter}

\title{Analysis method for detecting topological defect dark matter with a global magnetometer network}

\author[HIM]{Hector~Masia-Roig\corref{cor1}}
\ead{hemasiar@uni-mainz.de}
\author[HIM]{Joseph~A.~Smiga\corref{cor1}\corref{cor1}}
\ead{jsmiga@uni-mainz.de}
\author[HIM,Berkeley,LBNL]{Dmitry~Budker}
\author[Berkeley]{Vincent~Dumont}
\author[Fribourg]{Zoran~Grujic}
\author[IBS,KAIST]{Dongok~Kim}
\author[Hayward]{Derek~F.~Jackson~Kimball}
\author[Fribourg]{Victor~Lebedev}
\author[Hayward]{Madeline~Monroy}
\author[Jagiellonian]{Szymon~Pustelny}
\author[Fribourg,Leibniz]{Theo~Scholtes\fnref{TS}}
\author[Oberlin]{Perrin~C.~Segura}
\author[IBS,KAIST]{Yannis~K.~Semertzidis}
\author[IBS]{Yun~Chang~Shin}
\author[Oberlin]{Jason~E.~Stalnaker}
\author[Bucknell]{Ibrahim~Sulai}
\author[Fribourg]{Antoine~Weis}
\author[HIM]{Arne~Wickenbrock}

\cortext[cor1]{Corresponding author}
\fntext[TS]{Current address is Leibniz Institute of Photonic Technology. }

\address[HIM]{Helmholtz Institut Mainz, Johannes Gutenberg-Universit\"at, 55099 Mainz, Germany}
\address[Berkeley]{Department of Physics, University of California, Berkeley, CA 94720-7300, USA}
\address[LBNL]{Nuclear Science Division, Lawrence Berkeley National Laboratory, Berkeley, CA 94720, USA}
\address[Fribourg]{Physics Department, University of Fribourg, Chemin du Mus\'ee 3, CH-1700 Fribourg, Switzerland}
\address[IBS]{Center for Axion and Precision Physics Research, IBS, Daejeon 34051, Republic of Korea}
\address[KAIST]{Department of Physics, KAIST, Daejeon 34141, Republic of Korea}
\address[Hayward]{Department of Physics, California State University --- East Bay, Hayward, CA 94542-3084, USA}
\address[Jagiellonian]{Institute of Physics, Jagiellonian University, prof. Stanislawa Lojasiewicza 11, 30-348, Krak\'ow, Poland}
\address[Leibniz]{Leibniz Institute of Photonic Technology, Albert-Einstein-Straße 9, D-07745 Jena, Germany}
\address[Oberlin]{Department of Physics and Astronomy, Oberlin College, Oberlin, OH 44074, USA}
\address[Bucknell]{Department of Physics \& Astronomy, One Dent Drive, Bucknell University, Lewisburg, Pennsylvania 17837, USA}

\begin{abstract}
The Global Network of Optical Magnetometers for Exotic physics searches (GNOME) is a network of time-synchronized, geographically separated, optically pumped atomic magnetometers that is being used to search for correlated transient signals heralding exotic physics. GNOME is sensitive to exotic couplings of atomic spins to certain classes of dark matter candidates, such as axions. This work presents a data analysis procedure to search for axion dark matter in the form of topological defects: specifically, walls separating domains of discrete degenerate vacua in the axion field. An axion domain wall crossing the Earth creates a distinctive signal pattern in the network that can be distinguished from random noise. The reliability of the analysis procedure and the sensitivity of the GNOME to domain-wall crossings is studied using simulated data.
\end{abstract}

\end{frontmatter}

\section{Introduction}

The compelling evidence for dark matter~\cite{gorenstein_astronomical_2014} has inspired various theories to explain its nature~\cite{feng_dark_2010, Bergstrom2009}. Many of these theories propose new particles as dark matter candidates~\cite{feng_dark_2010, bertone_particle_2005}, and various experiments have been designed to search for these particles~\cite{Moriyama2008, safronova_search_2018, MarrodanUndagoitia2015, graham_experimental_2015, stadnik_searching_2015}. A well-motivated class of plausible dark matter constituents are axions and axion-like particles~\cite{duffy_axions_2009, ringwald2014axions}. The canonical QCD~axion was originally introduced to solve the strong-CP problem~\cite{peccei_cp_1977}, and variants of this idea have surfaced, for example, in string theory~\cite{arvanitaki2010string} and in solutions to the hierarchy problem~\cite{graham_cosmological_2015}. Hereafter, ``axion'' will refer to any axion-like particle and not only the canonical axion (which possesses particular constraints on the mass-coupling relationship).

Axions may form topological defects such as domain walls~\cite{sikivie_axions_1982, Kawasaki2015} or composite objects such as axion stars due to self-interactions~\cite{coleman_q-balls_1985, kusenko_q_2001, jackson_kimball_searching_2018, banerjee_relaxion_2019}. In particular, axion domain walls form between spatial domains wherein the axion field is centered around discrete vacua --- so the transition between these states must include field values that are not locally vacuum states. Axion domain walls are formed during a phase transition as the universe cools through expansion~\cite{pospelov_detecting_2013}. If the phase transition occurred after inflation, domain walls may continue to exist today; otherwise inflation would have pushed other domains outside of the observable universe. The domain walls may contain a substantial amount of energy, which may explain some component of dark matter~\cite{Kawasaki2015} and possibly dark energy~\cite{Friedland2003}. If the axion domain walls are a component of dark matter, it is reasonable to assume that they are virialized in the galaxy according to the standard halo model (SHM) with velocity dispersion of $\approx$290~km/s~\cite{freese_colloquium_2013, derevianko_hunting_2014, roberts_search_2017}. 
In this study, an analysis method is developed to search for axion domain walls using a global network of optical magnetometers, though the methods discussed in this paper could be applied to search for other objects such as axion stars.

The axion field can couple to ordinary matter in a variety of ways, as long as such interactions are not forbidden by fundamental laws of physics. For example, fermion spins may couple to the gradient of the axion field~\cite{pospelov_detecting_2013}. If fermionic matter crosses a region with an axion field gradient, such as a domain wall, it would experience a transient spin-dependent energy shift. This energy shift would appear as an effective magnetic field in atomic magnetometers which measure the energy-level splitting of different spin states.

To search for such transient spin-dependent effects, optical atomic magnetometers~\cite{Budker2007} were set up around the Earth to form the Global Network of Optical Magnetometers for Exotic physics searches (GNOME)~\cite{afach_characterization_2018, pustelny_global_2013}. At the core of each GNOME magnetometer is a vapor cell containing a gas of spin-polarized atoms. The atomic vapor cells are mounted within multi-layer magnetic shields that isolate them from external magnetic perturbations while retaining sensitivity to exotic fields causing spin-dependent energy shifts~\cite{Kimball2016}. Based on the experimental configuration, each magnetometer is sensitive to fields along a particular spatial axis and relatively insensitive to fields in the plane perpendicular to the sensitive axis. Each magnetometer has a characteristic bandwidth, typically $\approx 100$~Hz. There are additional sensors (e.g., accelerometers, gyroscopes, unshielded magnetometers, laser diagnostics) to monitor data quality. Under typical operating conditions, individual GNOME magnetometers experience occasional periods of poor-quality data which are flagged by these additional sensors. Furthermore, there are down times during which the magnetometers are off and no data are collected. The position, alignments of sensitive axes, and average noise background of nine of the magnetometers are shown in Table~\ref{table: MagInfo}. The noise background of each magnetometer is estimated by the average standard deviation of 30~min pre-processed data segments from December~2017. For further technical details on characteristics of the GNOME, see Ref.~\cite{afach_characterization_2018}.

\begin{table}[ht]
\centering
\caption{Characteristics of the GNOME sensors used for the simulated data. The positions, orientation of the sensitive axes, and noise are listed. The noise is the standard deviation of the Gaussian-distributed background used in the simulations.}
\begin{tabular}{ l D{.}{.}{8} D{.}{.}{7} D{.}{.}{0} D{.}{.}{0} D{.}{.}{2}  } 
\hline
\hline
          &                  \multicolumn{2}{c}{Location}                 &           \multicolumn{2}{c}{Orientation}           & \multicolumn{1}{l}{Noise} \\
Station   &  \multicolumn{1}{l}{Longitude} &  \multicolumn{1}{l}{Latitude} &   \multicolumn{1}{l}{Az} & \multicolumn{1}{l}{Alt} & \multicolumn{1}{l}{(pT)} \\ 
\hline 
Beijing   & 116.1868\textrm{\textdegree~E} & 40.2457\textrm{\textdegree~N} & +251\textrm{\textdegree} &   0\textrm{\textdegree} &  10.4 \\
Berkeley  & 122.2570\textrm{\textdegree~W} & 37.8723\textrm{\textdegree~N} &    0\textrm{\textdegree} & +90\textrm{\textdegree} &  14.5 \\
Daejeon   & 127.3987\textrm{\textdegree~E} & 36.3909\textrm{\textdegree~N} &    0\textrm{\textdegree} & +90\textrm{\textdegree} & 116   \\ 
Fribourg  &   7.1581\textrm{\textdegree~E} & 46.7930\textrm{\textdegree~N} & +190\textrm{\textdegree} &   0\textrm{\textdegree} &  12.6 \\ 
Hayward   & 122.0539\textrm{\textdegree~W} & 37.6564\textrm{\textdegree~N} &    0\textrm{\textdegree} & -90\textrm{\textdegree} &  14.3 \\ 
Hefei     & 117.2526\textrm{\textdegree~E} & 31.8429\textrm{\textdegree~N} &  +90\textrm{\textdegree} &   0\textrm{\textdegree} &  12.0 \\ 
Krakow    &  19.9048\textrm{\textdegree~E} & 50.0289\textrm{\textdegree~N} &  +45\textrm{\textdegree} &   0\textrm{\textdegree} &  15.6 \\ 
Lewisburg &  76.8825\textrm{\textdegree~W} & 40.9557\textrm{\textdegree~N} &    0\textrm{\textdegree} & +90\textrm{\textdegree} &  54.5 \\  
Mainz     &   8.2354\textrm{\textdegree~E} & 49.9915\textrm{\textdegree~N} &    0\textrm{\textdegree} & -90\textrm{\textdegree} &   6.8\\
\hline
\hline
\end{tabular}
\label{table: MagInfo}
\end{table}

If the Earth encounters a domain wall, a distinctive signal pattern is imprinted in the network. Signals would appear at each station at particular times and with particular amplitudes. The pattern is determined by the relative velocity between the Earth and the domain wall as well as the orientation of the sensitive axes of the magnetometers. These distinctive signal patterns are used to distinguish potential domain-wall crossing events from random noise. In the event of a discovery, signal characteristics can be used to extract information about the axion domain wall. For example, the physical thickness of the domain wall is inversely proportional to the axion mass~\cite{pospelov_detecting_2013}.

This paper describes an analysis algorithm to search for signal patterns in the GNOME data that are consistent with domain-wall crossing events and quantify their statistical significance. Additionally, a definition of network sensitivity is established that characterizes the properties of domain-wall signals observable by GNOME.

Before discussing the details of the analysis methods, a geometrical interpretation of the principles of the analysis procedure is introduced in Sec.~\ref{sec:geomPic}. The analysis procedure follows several steps that are described in detail in Sec.~\ref{sec:analysis}. The data are first binned and filtered to optimize the detection potential of the network. Then the processed data are analyzed to search for correlated signals matching the predicted pattern associated with the Earth crossing a domain wall. The magnetometers' data are time-shifted according to the expected delays. The most likely effective field vector associated with a potential domain wall is calculated at each time, accounting for the directional sensitivity of the sensors. Consistency between the expected and observed signals in the network is assessed to determine if the deviation between the observed and expected signal patterns can be explained by random noise. The statistical significance of a potential domain-wall crossing event is assessed according to its signal-to-noise ratio. Thresholds used to evaluate both the consistency with a domain-wall signal pattern and the statistical significance of the event are determined by studying false-positive and false-negative rates~\cite{panelli2019applying}. This analysis procedure is shown to be sensitive to domain-wall crossing events characterized by a particular range of parameters as discussed in Sec.~\ref{sec:netSensitivity}. The analysis algorithm is tested with simulated data, as described in Sec.~\ref{sec:testingMethods}. Finally, concluding remarks are given in Sec.~\ref{sec:conclusions}.

\section{Geometrical picture}\label{sec:geomPic}

A geometric viewpoint of the measurements is used to describe the analysis procedure. The magnetometer network measures the signals $\{s_i\}$ from a domain-wall crossing event in $n$ magnetometers, where $s_i$ corresponds to the amplitude measured at the $i^\text{th}$ magnetometer. A single measurement in the network can be expressed as an $n$-dimensional vector $\boldsymbol{s}$. The measurements, $\boldsymbol{s}$, have a corresponding uncertainty that can be expressed in terms of the covariance matrix $\Sigma_s$. Since the magnetometers have uncorrelated noise, $\Sigma_s$ is diagonal with entries corresponding to the respective variance in the magnetometer signals. For statistical considerations of significance, it helps to describe the measurements in terms of signal-to-noise ratios. The abstract vector space of all possible measurement vectors $\boldsymbol{s}$ can be rescaled by the noise, so that a point $\boldsymbol{u} \mapsto \boldsymbol{\tilde{u}} \equiv \Sigma_s^{-1/2} \boldsymbol{u}$, where $\Sigma_s^{-1/2}$ is the matrix square-root\footnote{Specifically, $\left( \Sigma_s^{-1/2} \right)^T \Sigma_s^{-1/2} = \Sigma_s^{-1}$. Existence of this matrix follows from the fact that the covariance matrix is positive definite. In this particular case, $\Sigma_s^{-1/2} = \text{diag}\{\sigma_i^{-1}\}$, where $\sigma_i$ is the noise of the $i^\text{th}$ magnetometer expressed as the standard deviation.} of $\Sigma_s^{-1}$. In the rescaled coordinates, each component of the vector $\boldsymbol{\tilde{s}}$ corresponds to the signal-to-noise ratio for some sensor.

The effective field associated with a domain-wall crossing event can be described using a three-dimensional vector $\boldsymbol{m}$ normal to the plane of the domain wall. This effective field vector will be referred to as the ``$m$-vector.'' For GNOME, the $m$-vector describes an effective magnetic field value due to coupling between atomic spins and an axion field. The strength of the signal is proportional to the norm $\lVert\boldsymbol{m}\rVert$. Note that there is some ambiguity since $\boldsymbol{m}$ can be either in the same or opposite direction to the relative velocity $\boldsymbol{v}$ between the domain wall and the Earth. One can relate $\boldsymbol{m}$ to the observed signal $\boldsymbol{s}$ with the linear equation $D\boldsymbol{m} \approx \boldsymbol{s}$, where $D$ is a $n \times 3$ matrix whose rows represent the sensitive direction of the magnetometers, adjusting for the interaction of an axion field with the particular atomic species used in each magnetometer~\cite{kimball2015nuclear}. Note that, in the event of a real domain-wall crossing signal, equality will not quite hold due to measurement uncertainty. To distinguish the measured amplitudes $\boldsymbol{s}$ from the expected observations from an $m$-vector $\boldsymbol{m}$, $\mu\equiv D\boldsymbol{m}$ is introduced. According to this linear equation,\footnote{It is possible to include non-linear effects, such as Earth's rotation and non-linear responses in the sensors, but these will not be considered here, because they are expected to be negligible. For sensors on Earth with domain walls traveling at $3\times 10^5$~m/s, the effects of the Earth's rotation will attenuate a signal by about 0.3\,\%.} all possible domain-walls signals are contained in a three-dimensional subspace spanned by the columns of $D$. The points in this subspace can be expressed as either the three-dimensional vector $\boldsymbol{m}$ or its corresponding point in the $n$-dimensional measurement space, $\mu=D\boldsymbol{m}$. 

\begin{figure}[ht]
	\centering
	\includegraphics[width=3.2in]{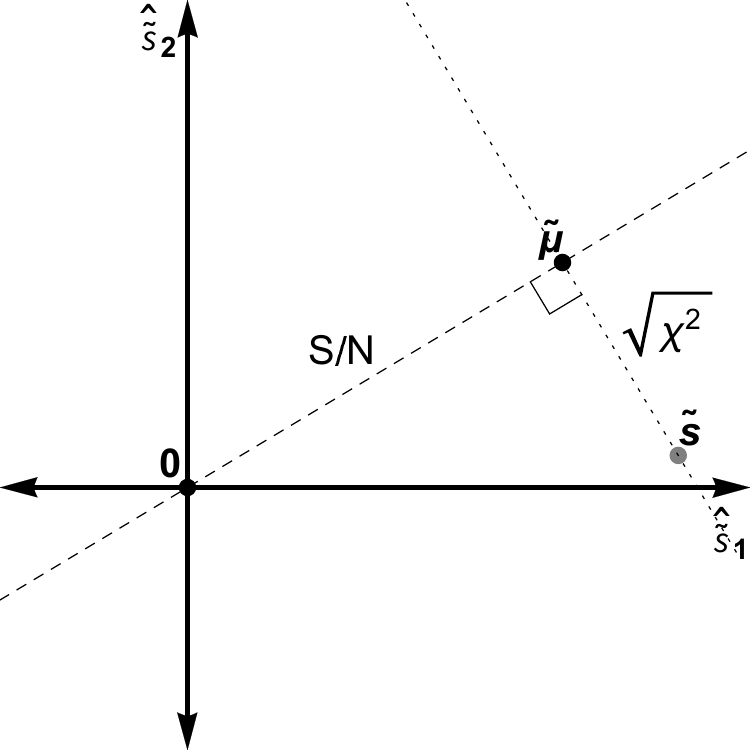}
	\caption{Geometric view of consistency check focusing on the plane in rescaled measurement space spanned by two sensors. A measured signal $\boldsymbol{\tilde{s}}$ shown with the expected amplitudes $\boldsymbol{\tilde{\mu}}$. The dashed line represents the subspace of possible domain wall measurements, while the dotted line represents the space of measurements that would yield the same ``expected'' signal. The degree of statistical agreement between $\boldsymbol{\tilde{s}}$ and $\boldsymbol{\tilde{\mu}}$ scales with their distance $\sqrt{\chi^2}$. The signal-to-noise is given by the magnitude of $\boldsymbol{\tilde{\mu}}$.}
	\label{fig:geometryView}
\end{figure}

A key parameter used to test the consistency of an observed signal $\boldsymbol{s}$ with that expected for a domain-wall crossing is the $\chi^2$. Given an expected domain wall effective field vector $\boldsymbol{m}_0$, the $\chi^2$ is defined as
\begin{equation}\label{eq:chi2}
\chi^2 = (\boldsymbol{s} - D\boldsymbol{m}_0)^T \Sigma_s^{-1} (\boldsymbol{s} - D\boldsymbol{m}_0)\ .
\end{equation}
For the distance $\boldsymbol{\Delta s}\equiv \boldsymbol{s}-\boldsymbol{\mu}_0$ (for $\mu_0\equiv D\boldsymbol{m}_0$), $\chi^2 = \boldsymbol{\Delta s}^T \Sigma_s^{-1} \boldsymbol{\Delta s} = \lVert \boldsymbol{\Delta \tilde{s}} \rVert^2$. Thus, the $\chi^2$ is the square of the distance from the set of measurements to the expected value in rescaled measurement space. Minimizing the $\chi^2$ is the same as finding the closest point between a measurement and a point $\boldsymbol{\tilde{\mu}}$ in the 3-dimensional subspace, which can be accomplished via a projection. A simplified two-dimensional pictorial model is shown in Fig.~\ref{fig:geometryView}.

In the GNOME analysis procedure, the geometric picture provided in this section serves as a means of visualizing the data. In the rescaled measurement space, distances represent the degree of statistical agreement and measurements corresponding to domain-wall crossing events exist in a three-dimensional linear subspace. Values $\boldsymbol{s}$ in measurement space are constructed by sampling values from each magnetometer at some time accounting for expected delays. The delays are estimated by selecting a particular domain-wall crossing velocity. As a result, a measurement $\boldsymbol{s}$ can be generated for any given time and velocity, since each velocity results in a different set of delays. The direction of the velocity should be in agreement with the direction of the calculated $\boldsymbol{m}$. 


\section{Analysis procedure}\label{sec:analysis}

The analysis procedure presented here is designed to search the GNOME data for domain-wall crossing events. These events are modeled as a plane of finite thickness that travels through the Earth at a constant velocity. For a given plane orientation and speed,\footnote{For an ideal plane, only the velocity perpendicular to the plane is observable. Thus, the velocity is entirely described by the speed and normal direction of the wall.} the signal pattern in the sensor network can be predicted. Assuming a linear coupling between the axion field gradient and fermion spins (i.e., of the form $J^\mu \partial_\mu a$ for $J^\mu$ related to the fermion spin and $a$ being the axion field~\cite{pospelov_detecting_2013}), a transient pulse will appear in the measured magnetic field data as the domain wall crosses the Earth \cite{pospelov_detecting_2013}. The transient pulse amplitude observed by an individual GNOME sensor is also affected by the specific axion-field coupling to that atomic species~\cite{kimball2015nuclear} and the angle between the axion-field gradient and the sensitive axis of the sensor~\cite{afach_characterization_2018}.

\begin{figure}[ht]
	\centering
	\includegraphics[width=3.2in]{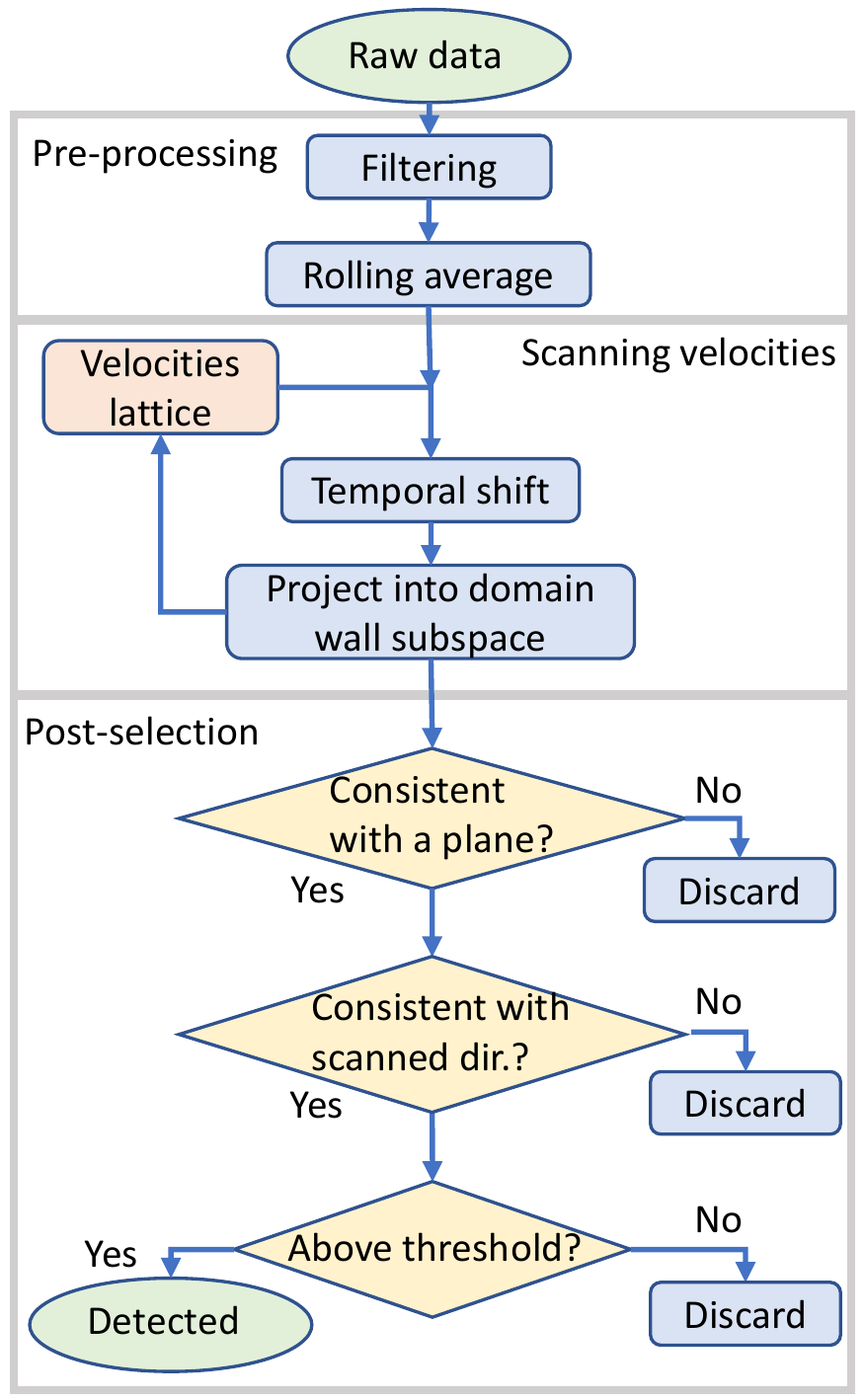}
	\caption{Flowchart describing the analysis algorithm used to detect domain-wall crossing events in the GNOME data. After pre-processing, the data are aligned in time according to a velocity scanning grid (see Sec.~\ref{sec:ScanSky}). Comparing the measured and expected signals, different statistical parameters are extracted to quantify the agreement (see Sec.~\ref{sec:projection}). Thresholds are applied to these statistical parameters to filter out plausible domain-wall crossing signals (see Sec.~\ref{sec:thresholds}). }
	\label{fig:FlowChart}
\end{figure}

The analysis procedure is composed of three steps designed to find domain-wall events. First, in the pre-processing stage, the raw data are filtered and a rolling average is applied in order to enhance the detection capabilities of the network. Second, in the velocity-scanning stage, the data from the individual magnetometers are time-shifted according to different domain-wall velocities. This ensures that the transient signals corresponding to a domain-wall crossing  appear simultaneously in all magnetometers. Third, in the post-selection stage, each network measurement is characterized by three parameters: direction, magnitude, and consistency between the observed signal pattern and the expected signal for a domain-wall crossing. If an event passes a set of thresholds applied to these three parameters, it will be considered  statistically significant (see Sec.~\ref{sec:projection}). A basic flowchart of the procedure can be seen in Fig.~\ref{fig:FlowChart}.

\subsection{Data pre-processing}
In order to optimize for domain-wall search, the data are pre-processed through filtering and a rolling average is taken. Filters are used to remove long-term drifts as well as noisy frequency bands, e.g., the power-line frequency~\cite{afach_characterization_2018}. After filtering, we perform a rolling average of the data over time $T$. Averaging the data enhances the signal-to-noise ratio for a certain signal duration and avoids complications arising from different magnetometers having different bandwidths. However, filtering and averaging data will also attenuate and modify the shape of the signal lineshape. A detailed analysis of the effects of filtering and averaging on the data is given in \ref{app:filterBin}.

The filters attenuate frequency bands containing known noise sources, however some non-Gaussian noise from unidentified sources may remain. Therefore, the noise is determined after the pre-processing steps. The uncertainty at a given time is estimated by the standard deviation of the data around that time, taken in periods of $T$ to prevent correlations due to the rolling average. In order to minimize the effects of a signal in the estimation of the noise, outliers are removed from the calculation of the standard deviation.

\subsection{Scanning over velocities}\label{sec:ScanSky}
After the pre-processing stage, the data are time-shifted so that a domain-wall signal would appear at all magnetometers at the same time. This is possible because, for a given relative velocity between a domain wall and the Earth, the magnetometer signals appear in a predictable pattern. 

\begin{figure}[ht]
	\centering
	\includegraphics[width=6.4in]{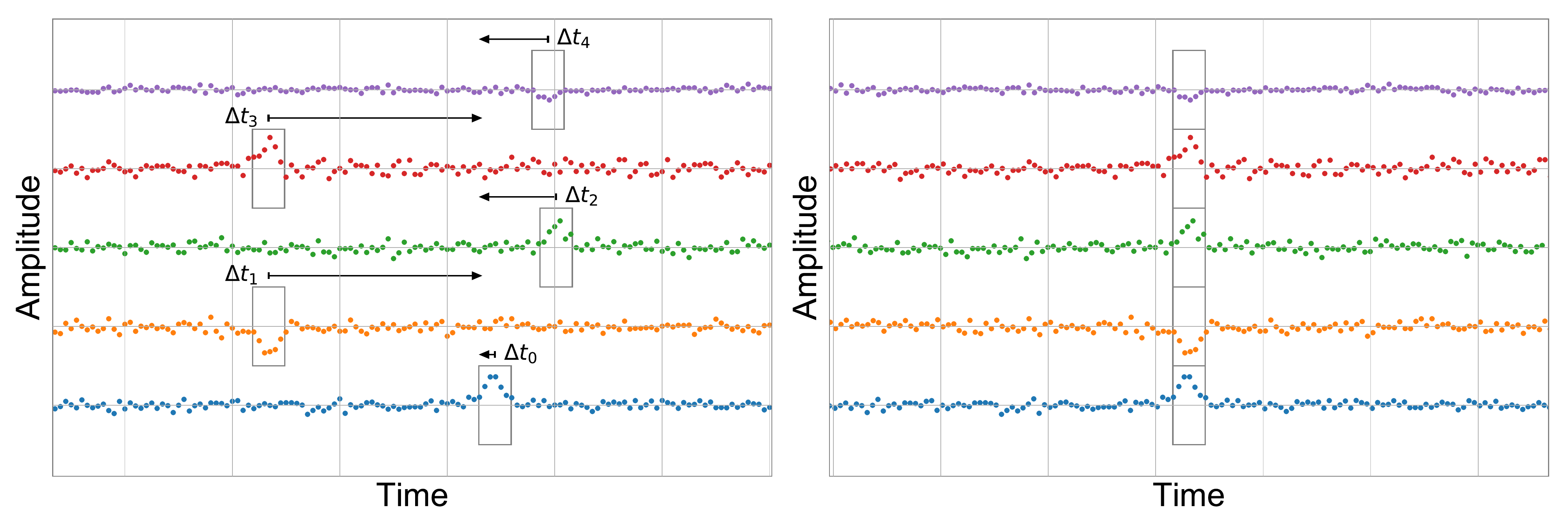}
	\caption{Simulated GNOME data featuring a domain wall signal before time-shifting (left) and after (right). The $\Delta t_i$ for the different stations are determined by their geographical location and the velocity of the domain wall. The different amplitudes are determined by the orientation of the sensitive axes of the detectors relative to the domain wall velocity.}
	\label{fig:delay}
\end{figure}

The sensors in the network are located at different positions, $\{ \boldsymbol{x}_i \}$ on the surface of the Earth. A domain wall with speed $\lVert\boldsymbol{v}\rVert$ in direction $\hat{\boldsymbol{v}}$ crossing the Earth is observed by different sensors at times $\{t_i\}$.  The time difference from when a wall passes two locations can be expressed as
\[
\Delta t_{i}=(\boldsymbol{x}_\text{i}-\boldsymbol{x}_0)\cdot \frac{\boldsymbol{v}}{\lVert\boldsymbol{v}\rVert^2}\ ,
\]
where the sensor at $\boldsymbol{x}_0$ is used as a reference. The time at each data point is shifted according to $\Delta t_{i}$ to align all the signals. The delays $\Delta t_{i}$ are calculated in intervals of $T/2$. Then the corresponding points are extracted from the rolling averaged data. After this operation, an aligned set of measurements calculated with overlapping averaging windows is obtained. A graphical representation of the time shifting operation can be seen in Fig.~\ref{fig:delay}.

Earth-based sensors are in a noninertial (rotating) reference frame. For Earth's radius ($\approx 6.4\times10^6$~m), rotation period (1~day), and a domain-wall velocity of $v\approx3\times 10^5$~m/s, according to numerical estimates, the additional signal delay due to the Earth's rotation would be $\Delta t \lesssim 33$~ms. This effect is significant compared to the sensor bandwidth (expected to be $\approx 250$~Hz) and should be corrected. To first order, the sensors are all moving at a constant velocity tangential to the Earth's surface, though only the component that is normal to the wall is observable. Including this correction,
\begin{equation}
    \Delta t_i = \boldsymbol{\Delta x_i}\cdot \frac{\boldsymbol{v}}{\lVert \boldsymbol{v} \rVert^2 - \boldsymbol{\delta v_i}\cdot \boldsymbol{v}}\ ,
    \label{eq:delay}
\end{equation}
where $\boldsymbol{\delta v_i}$ is the tangential velocity of the $i^\text{th}$ sensor at the crossing time (when the wall crosses the center of the Earth). The first-order correction reduces the relative time error to $\Delta t \lesssim 0.05$~ms; well below the bandwidth of the GNOME magnetometers.

Assuming the SHM, the distribution of domain wall velocities can be predicted.  Within this model, the dark matter structures are virialized in the galaxy. This means that the domain-wall velocity distribution is isotropic and quasi-Maxwellian\footnote{It is quasi-Maxwellian as opposed to Maxwellian due to the cut-off at the galactic escape velocity and the relative velocity of the Earth.} with dispersion $\sigma_v\approx290$~km~s$^{-1}$ and a cut-off above the galactic escape velocity of $v_\text{esc}\approx550$~km~s$^{-1}$~\cite{roberts_search_2017}. The Earth moves through the dark matter halo with apparent velocity towards the Cygnus constellation. A range of speeds and relative angles with respect to the Earth movement are selected in the analysis so that 95\,\% of the expected velocities are observable.

The scanning step size is estimated by considering two antipodal magnetometers. From Eq.~\eqref{eq:delay} the changes in the delay time with respect to variation in the speed can be estimated. However, the delay is also dependent on the direction of the wall. In order to give an upper bound, the direction giving the largest variation of the delay is chosen. In addition, it is required that the maximum delay change must be smaller than half the bin size, $T/2$, so the signal remains in the same bin. The speed range given by the SHM is scanned in steps of
\begin{equation}
 \delta v \leq \frac{T v^2}{4 R_\oplus}\ ,
\label{eq:VelStep}
\end{equation}
where $R_\oplus$ is the radius of the Earth. The same procedure can be followed to establish a scanning step for the angles. The step is given by
\begin{equation}
 \delta \theta \leq \frac{T v}{4 R_\oplus}\ .
\label{eq:AngStep}
\end{equation}
For a given speed, a lattice on the celestial hemisphere should have a point within every circle whose diameter spans an arc of $\delta\theta$. Note that the scanning step size is dependent on the speed.

To determine the lattice of directions, a set of points evenly distributed on the sphere are needed. One wants to guarantee that any circle whose radius on the sphere is given by $\delta\theta$ [Eq.~\eqref{eq:AngStep}] contains at least one scanned direction. A roughly even distribution of points on the sphere is generated using the Fibonacci lattice method, with the number of points based on the step size (see, e.g., Ref.~\cite{swinbank_fibonacci_2006} for a description). Briefly, the Fibonacci lattice method is a means of generating a sequence of points that covers a surface. In this case, each sequential point has an azimuthal angle that increments by a factor of $2\pi\varphi$, where $\varphi\equiv \frac{1+\sqrt{5}}{2}$ is the golden-ratio, while the polar angle is incremented such that the points are evenly distributed between the poles.

For each velocity, an abstract ``measurement space'' is constructed as described in Sec.~\ref{sec:geomPic}. After adjusting for delays, the amplitudes measured at a given time belong to the same event. The events are represented as a vector in an $n$-dimensional space, where $n$ is the number of magnetometers. However, measured events corresponding to a domain-wall crossing must lie in a 3-dimensional subspace of the measurement space. The application of the mathematical tools presented in Sec.~\ref{sec:geomPic} to the time-shifted data is discussed in the following sections.



\subsection{Post-selection}
After the measurements are temporally aligned according to the scanned velocities, their agreement with a domain-wall crossing is assessed by comparing the expected domain-wall signal pattern with the observed pattern. In the geometrical picture, the measured event is projected to the domain-wall subspace in coordinates scaled by the noise of the magnetometers, and the distance between the measurement and projected value quantify the statistical agreement of the observation with an expected measurement. Three parameters are relevant to determine if a set of measurements is statistically significant: the $p$-value measuring the statistical agreement between the measured signals $\boldsymbol{s}$ and an expected domain-wall crossing signal $\boldsymbol{\mu}$, the angle between the scanned velocity $\boldsymbol{\hat{v}}$ and observed wall orientation $\boldsymbol{\hat{m}}$, and the signal-to-noise ratio of $\lVert\boldsymbol{m}\rVert$. 

\subsubsection{Project into subspace}\label{sec:projection}

After time-shifting the data for a given velocity, one obtains a measurement $\boldsymbol{s}$ at every time consisting of data from all active sensors. At each time, there is an expected domain-wall crossing signal $\boldsymbol{\mu}\equiv D\boldsymbol{m}$ that is the closest point in the subspace of domain-wall signals to $\boldsymbol{s}$ when using rescaled coordinates (as described in Sec.~\ref{sec:geomPic}). The $m$-vector $\boldsymbol{m}$ describes the effective field associated with the domain-wall crossing event. Thus for every scanned velocity, a ``most likely'' $m$-vector is found for the $\boldsymbol{s}$ at every time; i.e., the $m$-vector that would result in an expected signal that most closely reproduced the observed signal. In the next stages of the analysis it is determined whether $\boldsymbol{s}$ is in statistical agreement with the ``most likely'' domain-wall crossing event and cannot be explained by random noise.

One can assume that the amplitudes from the $n$~sensors $\{s_i\}$ (for $i=1,\ldots,n$) obey a linear equation with signal attenuation caused by misalignment between the magnetometers' sensitive directions $\{\hat{\boldsymbol{d}}_i\}$ and the effective magnetic field induced by the axion field:
\begin{equation}\label{eq:ampLinEq}
D\boldsymbol{m} = \boldsymbol{s}\quad \text{for}\ D\equiv \left[ 
\begin{matrix}
\hat{\boldsymbol{d}}_1^T \\
\hat{\boldsymbol{d}}_2^T \\
\vdots \\
\hat{\boldsymbol{d}}_n^T
\end{matrix}
\right],\ \boldsymbol{s} \equiv \left[
\begin{matrix}
s_1 \\
s_2 \\
\vdots \\
s_n
\end{matrix}
\right]\ ,
\end{equation}
where $\boldsymbol{m}$ is the three-dimensional $m$-vector whose norm represents the strength of the effective field and whose direction is normal to the domain wall. In general, the magnetometers are expected to experience different (though still linear) responses to an event due to, different couplings of the axion field to different atomic species \cite{kimball2015nuclear}. These effects can be included by multiplying the corresponding row in $D$ by the appropriate response factor.

\begin{figure}[ht]
    \centering
    \includegraphics[width=5.9in]{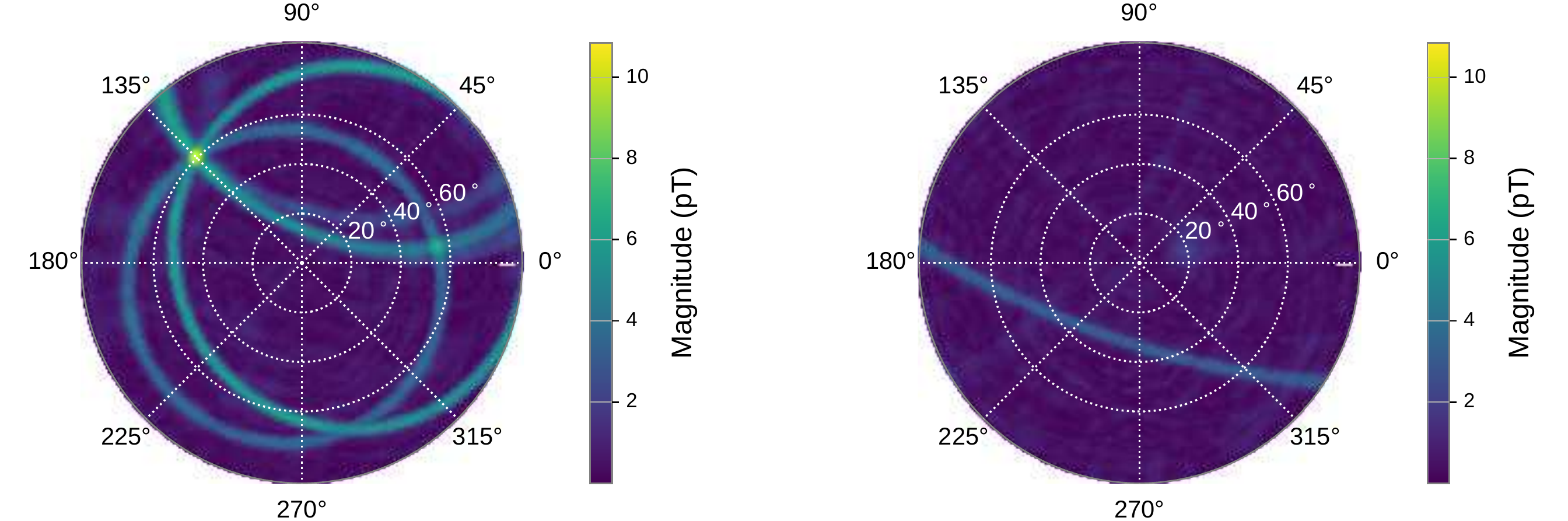}
    \caption{Illustration of the sigal magnitude proportional to  $\lVert\boldsymbol{m}\rVert$ found at different directions, $\hat{v}_\text{scan}$, across a single hemisphere. Left: a domain-wall crossing event is inserted in the data with effective field magnitude corresponding to 20~pT and direction at polar angle~$60^\circ$ and azimuthal angle~$135^\circ$. Right: the same domain-wall crossing event is inserted but the time delays at each magnetometer are randomized. The simulation is performed using the magnetometers' characteristics from Table~\ref{table: MagInfo}.}
    \label{fig:Skymap}
\end{figure}

As discussed in Sec.~\ref{sec:geomPic}, solving Eq.~\eqref{eq:ampLinEq} as a least-squares minimization problem --- given amplitudes $\boldsymbol{s}$ and covariance $\Sigma_s$ --- is equivalent to performing a fit/projection of $\boldsymbol{s}$ into the subspace spanned by the columns of $D$. The result is
\begin{equation}
\boldsymbol{m} = \Sigma_m D^T \Sigma_s^{-1} \boldsymbol{s}\quad \text{for}\quad \Sigma_m = (D^T\Sigma_s^{-1}D)^{-1}\ .
\label{eq:proyection}
\end{equation}
Scanning velocities produces different values for $\boldsymbol{s}$ at a given time, and therefore, different values for $\boldsymbol{m}$. A maximum on the norm of $\boldsymbol{m}$ is expected when the scanned velocity corresponds to the domain-wall crossing velocity present in the data, as can be seen in Fig.~\ref{fig:Skymap}.

An important statistical result from the fit is the $\chi^2$ [Eq.~\eqref{eq:chi2}], which describes the deviation between a measurement and expected signal pattern. Assuming that the noise in the measurements are normally distributed, the $\chi^2$ values are distributed according to the number of degrees of freedom ($\dim \boldsymbol{s}-\dim \boldsymbol{m}=n-3$). The $p$-value is given by the integrated right tail of this distribution starting from the measured $\chi^2$. The $p$-value corresponds to the probability that the residual between the expected and measured values can be explained by deviations due to Gaussian noise. 

\subsubsection{Thresholds}\label{sec:thresholds}
For each time and scanned velocity, a signal vector $\boldsymbol{s}$  and its corresponding $m$-vector and $p$-value are determined. Measurement vectors consistent with domain-wall crossings must be distinguished from signals originating from noise or systematic effects. This identification is accomplished by imposing thresholds on the $p$-value, the signal-to-noise ratio, and the direction of $\boldsymbol{m}$.

\begin{figure}[ht]
	\centering
	\includegraphics[width=5.9in]{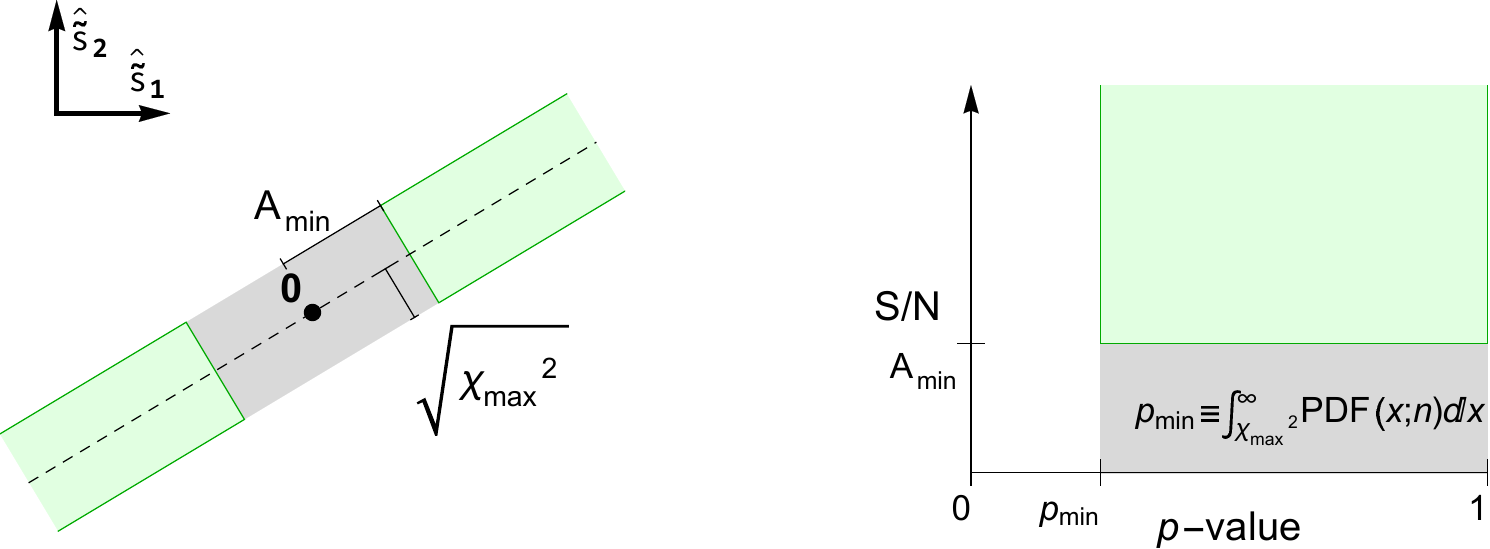}
	\caption{Geometric view of consistency check simplified in a lower dimension. The dashed line on the left image represents the 3-dimensional subspace of expected signals. A visualization of the thresholds where the green shaded region is accepted as a likely and significant signal. The signal-to-noise ratio threshold is $A_\text{min}$ while the $\chi^2$ threshold is $\chi^2_\text{max}$.}
	\label{fig:geoThreshold}
\end{figure}

\begin{figure}[ht]
    \centering
    \includegraphics[width=3.3in]{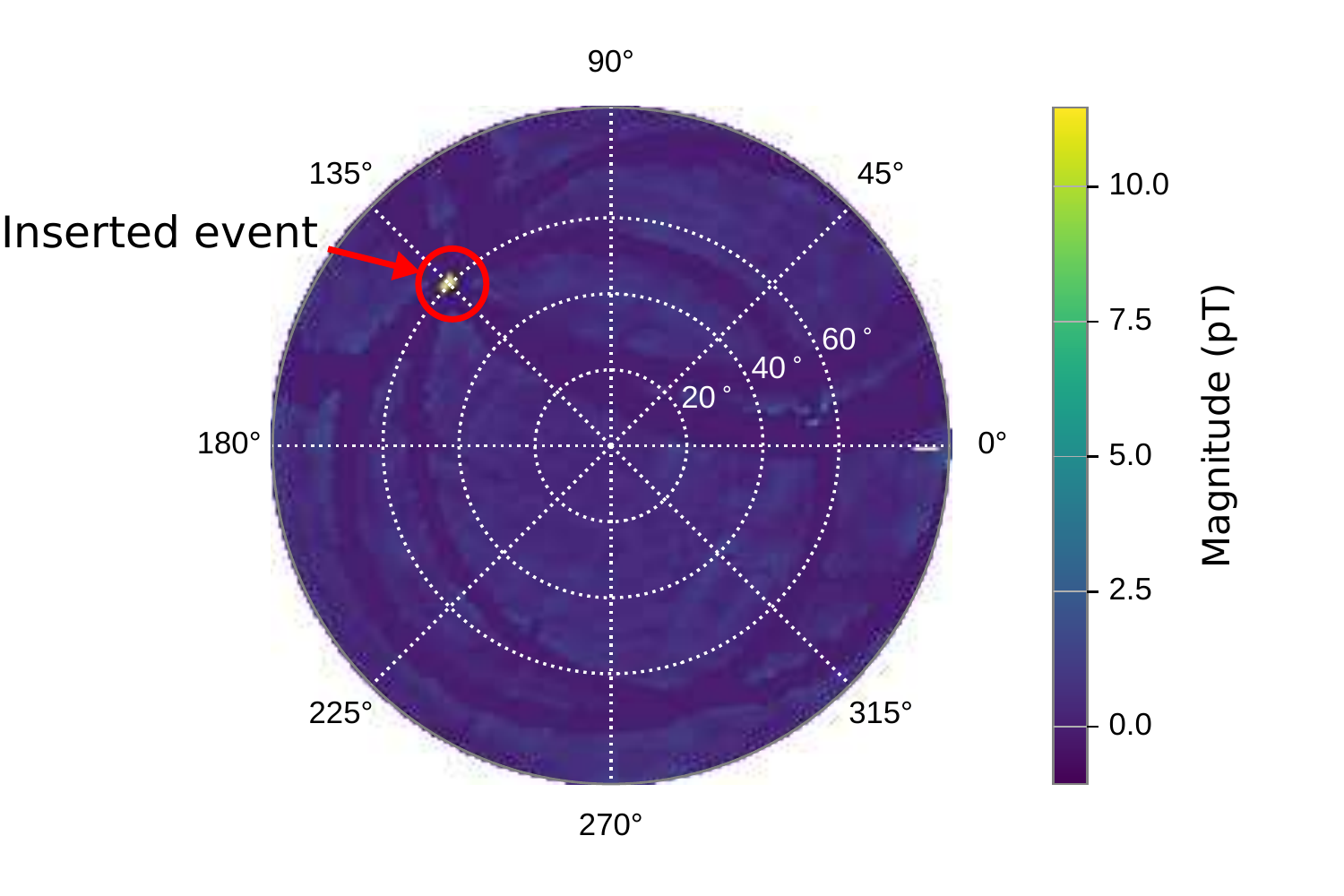}
    \caption{Illustration of the signal magnitude found at different velocities, $\hat{v}_\text{scan}$, across a single hemisphere. The speed is kept constant. The same data as Fig.~\ref{fig:Skymap} is used, however the requirement of a $p$-value greater than 0.05 is imposed. The high-magnitude points are now reduced to a point corresponding to the direction where the domain wall was injected.}
    \label{fig:SkymapPval}
\end{figure}

The agreement between the observed event and the domain-wall crossing model is quantified by the $p$-value. This is related to the distance from the measured point $\boldsymbol{s}$ to the subspace of domain walls, see Fig.~\ref{fig:geoThreshold}. If the $p$-value is small, the candidate event can be rejected because the deviation from the expected signal pattern is too large to be explained by uncertainty in the measurement. For instance, if two sensors have the same sensitive direction, then it is unlikely that they would report significant amplitudes with opposite sign. The $p$-value is a powerful tool for rejecting spurious spikes in signals from individual magnetometers, as can be seen in Fig.~\ref{fig:SkymapPval}. The magnitude reported could be large, however the $p$-value would be small because the other magnetometers would not feature a signal. The $p$-value threshold is chosen so that only 5\,\% of real events would be misidentified as noise. For Gaussian-distributed noise, this corresponds to a $p$-value threshold of $0.05$, meaning that only events with greater $p$-values are processed further. However, if the noise is more complex, the $p$-value corresponding to 5\,\% false-negatives has to be explicitly calculated, as shown in Sec.~\ref{sec:false-negative}.


The data from each magnetometer are time-shifted according to a discrete set of velocities (see Sec.~\ref{sec:ScanSky}). However, the direction $\hat{\boldsymbol{m}}$ is reconstructed independent of the scanning velocity, $\boldsymbol{v}_{scan}$. Therefore, the agreement between the scanned and reconstructed directions must be checked. If the angular difference between $\boldsymbol{v}_{scan}$ and $\boldsymbol{m}$ is found to be larger than the angular lattice spacing, from Eq.~\eqref{eq:AngStep}, the event is rejected; it is inconsistent with a domain-wall crossing because the velocity $\hat{\boldsymbol{v}}$ is not parallel to the axion field gradient $\hat{\boldsymbol{m}}$. 


After an event has passed the consistency checks, its significance has to be evaluated in terms of magnitude. The magnitude is given by the norm of the projection of $\boldsymbol{s}$ to the domain-wall subspace,
\begin{equation}\label{eq:magNorm}
    \lVert \boldsymbol{m} \rVert \pm \frac{1}{\lVert \boldsymbol{m} \rVert} \sqrt{\boldsymbol{m}^T \Sigma_m \boldsymbol{m}}\ ,
\end{equation}
where $\Sigma_m$ is the covariance matrix of the $m$-vector defined in Eq.~\eqref{eq:proyection}. The quotient of the norm and its uncertainty is the signal-to-noise ratio. Events featuring a large signal-to-noise ratio are less likely to be produced by noise. Since the noise in the network is not purely Gaussian, the specific signal-to-noise ratio needed to claim a detection is fixed by studying the data. For this, a data set not containing any sought signal but featuring the typical noise characteristics of the network is analyzed. The rate of events found is studied with respect to their signal-to-noise ratio. Then the probability of finding an event above certain signal-to-noise threshold is assessed. This is called false-positive analysis and a case with simulated data is evaluated in Sec.~\ref{sec:testingMethods}. The thresholds are visualized in Fig.~\ref{fig:geoThreshold}.


The rate of events found is expected to follow Poissonian statistics. Namely, the probability that one finds $n_f$ events over an interval of duration $t$, with an expected occurrence rate, $r$, is given by
\begin{equation}\label{eq:PoissonProb}
P(n_f;rt) = e^{-rt} \frac{(rt)^{n_f}}{n_f!}\ .
\end{equation}
For an interval $T_\text{samp}$ of data, if $n_f$ events are found, the upper-bound on the rate $r_0\geq r$ at a confidence level $C$ is given by solving
\begin{equation}\label{eq:rateBound}
C = \int_0^{r_0T_\text{samp}}P(n_f;x)dx\ .
\end{equation}
From Eq.~\eqref{eq:PoissonProb}, the probability of finding more than zero events over the course of a $t$-long run is then
\begin{equation}
    P_\text{FP} \leq 1-e^{-r_0t}\ .
\end{equation}
To reach $5\sigma$ significance for detection, the maximum probability for finding more than zero events must be $P_\text{FP}<5.7\times10^{-7}$, or 1 in 1.7~million, over the course of a data collection run. The signal-to-noise threshold for detection is chosen so the rate of events found is smaller than 1 in~1.7 million. An example with simulated data is given in Sec.~\ref{sec:testingMethods}. 

Note that if no domain-wall crossing event is found above the detection threshold, no detection can be claimed. The event found with the maximum signal-to-noise ratio defines the detection threshold of the network for the measured time interval.

\section{Network sensitivity}\label{sec:netSensitivity}
In order to define the detection capabilities of GNOME, a notion of sensitivity must be established. Defining the matrix $D$ and $m$-vector $\boldsymbol{m}$ as in Eq.~\eqref{eq:ampLinEq}, one can define a function $\mathcal{A}$ that takes the effective field vector $\boldsymbol{m}$, noise $\Sigma_s$, and $D$ and returns a collective signal-to-noise ratio.

The output of this function is compared to some threshold $\alpha$ for finding a domain-wall crossing event. For example, if $\mathcal{A}(\boldsymbol{m},\Sigma_s,D)\geq \alpha$ for the event, then a signal is found, otherwise it is missed. Thus, the exact definition of $\mathcal{A}$ is based on the analysis method. For the analysis described here, one finds signals by searching for instances in which the norm of the $m$-vector exceeds some multiple of its uncertainty. According to Eq.~\eqref{eq:magNorm}, one finds
\[
\mathcal{A}(\boldsymbol{m},\Sigma_s,D)=\frac{m}{\sqrt{\boldsymbol{\hat{m}}^T (D^T\Sigma_s^{-1}D)^{-1}\boldsymbol{\hat{m}} }}\ .
\]
Observe that when $\boldsymbol{m}$ is an eigenvector of $D^T\Sigma_s^{-1}D$, then $\mathcal{A}(\boldsymbol{m},\Sigma_s,D)=\lVert \Sigma_s^{-1/2}D\boldsymbol{\hat{m}} \rVert\equiv \mathcal{A}'(\Sigma_s^{-1/2}D\boldsymbol{\hat{m}})$.

The sensitivity of the system can be defined as the minimum signal needed to guarantee that the signal-to-noise is at least $\alpha$. The sensitivity in the direction $\hat{\boldsymbol{m}}$ is obtained by solving $\mathcal{A}(\beta_\alpha\boldsymbol{\hat{m}},\Sigma_s,D)=\alpha$ for $\beta_\alpha$:
\begin{equation}
    \beta_\alpha(\hat{\boldsymbol{m}}) \equiv \frac{\alpha}{\mathcal{A}(\hat{\boldsymbol{m}},\Sigma_s,D)} = \alpha\sqrt{\boldsymbol{\hat{m}}^T (D^T\Sigma_s^{-1}D)^{-1}\boldsymbol{\hat{m}}}\ ,
    \label{eq:sensitivity}
\end{equation}
since $\mathcal{A}$ is absolutely scalable --- i.e., $\mathcal{A}(\beta\hat{\boldsymbol{m}}) = |\beta| \mathcal{A}(\hat{\boldsymbol{m}})$. Thus, if $\beta_\alpha(\hat{\boldsymbol{m}})$ is large, then a large magnitude is needed to induce a measurable signal in the direction $\hat{\boldsymbol{m}}$. The signal-to-noise threshold will be $\alpha$, so a stricter, higher threshold results in a proportionally worse sensitivity.

\begin{figure}[ht]
    \centering
    \includegraphics[width=6.4in]{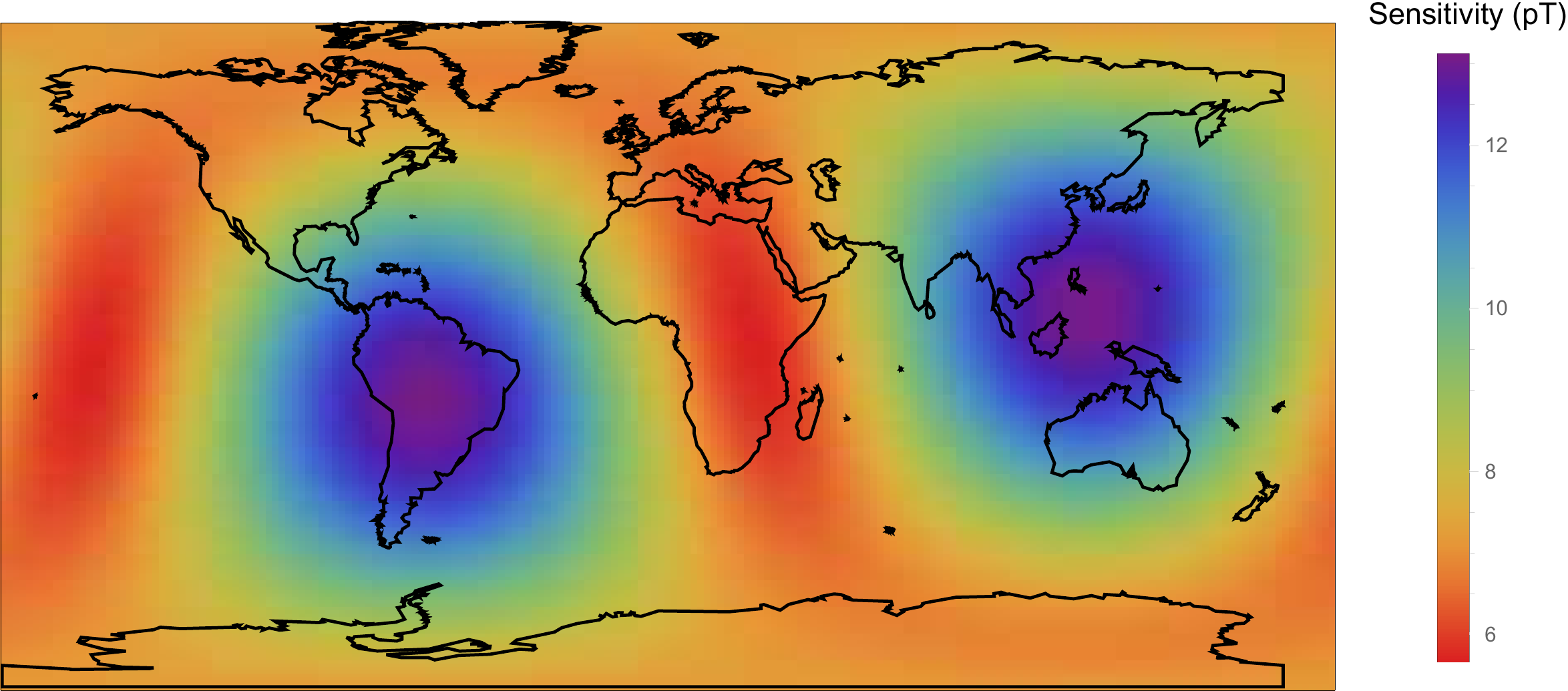}
    \caption{Directional sensitivity of the network according to the configuration used to generate simulated data (see Table~\ref{table: MagInfo}). The color overlayed on the map of the Earth is $\beta_1(\hat{\boldsymbol{m}})$ from Eq.~\eqref{eq:sensitivity}, where the position on the map corresponds to the first contact point of a domain wall on the Earth's surface.}
    \label{fig:avgSensMap}
\end{figure}

An example of the network sensitivity is plotted in Fig.~\ref{fig:avgSensMap} in geocentric coordinates for $\alpha=1$. The configuration of the sensors is described in Table~\ref{table: MagInfo}. A clear pattern can be observed where the network is more sensitive to certain directions. An ideal configuration would show an homogeneous sensitivity in all directions. Nevertheless, the network features a fairly uniform sensitivity, only varying by a factor of two between the best and worse direction.

To reduce $\beta_\alpha(\hat{\boldsymbol{m}})$ to a single number, one could assume a distribution of signals based on some model (e.g., the SHM) and take the weighted average of the sensitivity over the signal distribution. Alternatively, one could achieve a sensitivity bound by considering the worst-case scenario in which $\beta_\alpha$ is maximized. In this case, this is accomplished by finding the smallest eigenvalue of $D^T\Sigma^{-1}D$. Denote $\lambda_\text{min}$ as the smallest eigenvalue and $\hat{\boldsymbol{x}}_\text{min}$ as the corresponding eigenvector. Then the sensitivity in the worst direction is $\alpha/\lambda_\text{min}$ for the worst direction $\hat{\boldsymbol{x}}_\text{min}$. Note that the optimal orientation for adding an additional sensor to the network would be $\hat{\boldsymbol{x}}_\text{min}$ in any location. Additionally, filtering and binning will alter the sensitivity of the network to particular signal shapes (e.g., for signals with different widths). These effects are discussed in \ref{app:filterBin}.


\section{Testing analysis methods}\label{sec:testingMethods}

The previous sections present the analysis algorithm and the relevant statistical parameters to identify domain-wall crossings in the GNOME data. In this section, this analysis algorithm is tested with simulated data. The reliability of the algorithm is assessed based on the false-negatives and false-positive rates. False negatives occur when a domain-wall crossing is present but the algorithm fails to identify it, while false positives occur when noise is wrongly identified as a domain-wall crossing. 

\subsection{False-negative analysis}\label{sec:false-negative}

The proposed algorithm has to be able to identify signals which match the characteristics of a domain-wall crossing event occurring at any time in the data. The expected directions and speeds of crossings are described by a probability distribution based on the SHM (see Sec.~\ref{sec:ScanSky}). Though the magnitude of $\boldsymbol{m}$ and the duration of the domain-wall crossing can take any values, the range of observable values is limited by the sensitivity of the sensor network (see Sec.~\ref{sec:netSensitivity}). 

Twenty-minute-long simulated data segments with 512~Hz sampling rate are constructed with Gaussian-distributed noise according to Table~\ref{table: MagInfo}. A Lorentzian-shaped pulse is added into the data of each magnetometer according to the model of a domain-wall crossing event for a given velocity. The timing and amplitudes of the pulses are calculated based on Eq.~\eqref{eq:delay} and Eq.~\eqref{eq:ampLinEq}. The crossing time is defined to be the moment the domain wall crosses the center of the Earth; this fixes the relative delays. 

For the false-negative analysis, the crossing time and domain wall direction $\hat{\boldsymbol{v}}$ are randomized while the speed is kept constant. An effective field magnitude corresponding to 20~pT is chosen so that the signal amplitudes are clearly visible in the averaged data. A rolling average of the data is taken with averaging time of $T=1~\text{s}$, and a high-pass filter with a cut-off frequency of 1/300~Hz is applied to the data. Moreover, notch filters are applied corresponding to the electric network frequency for each station to include the effects of filtering on the signal.

\begin{figure}[ht]
    \centering
    \includegraphics[width=3.2in]{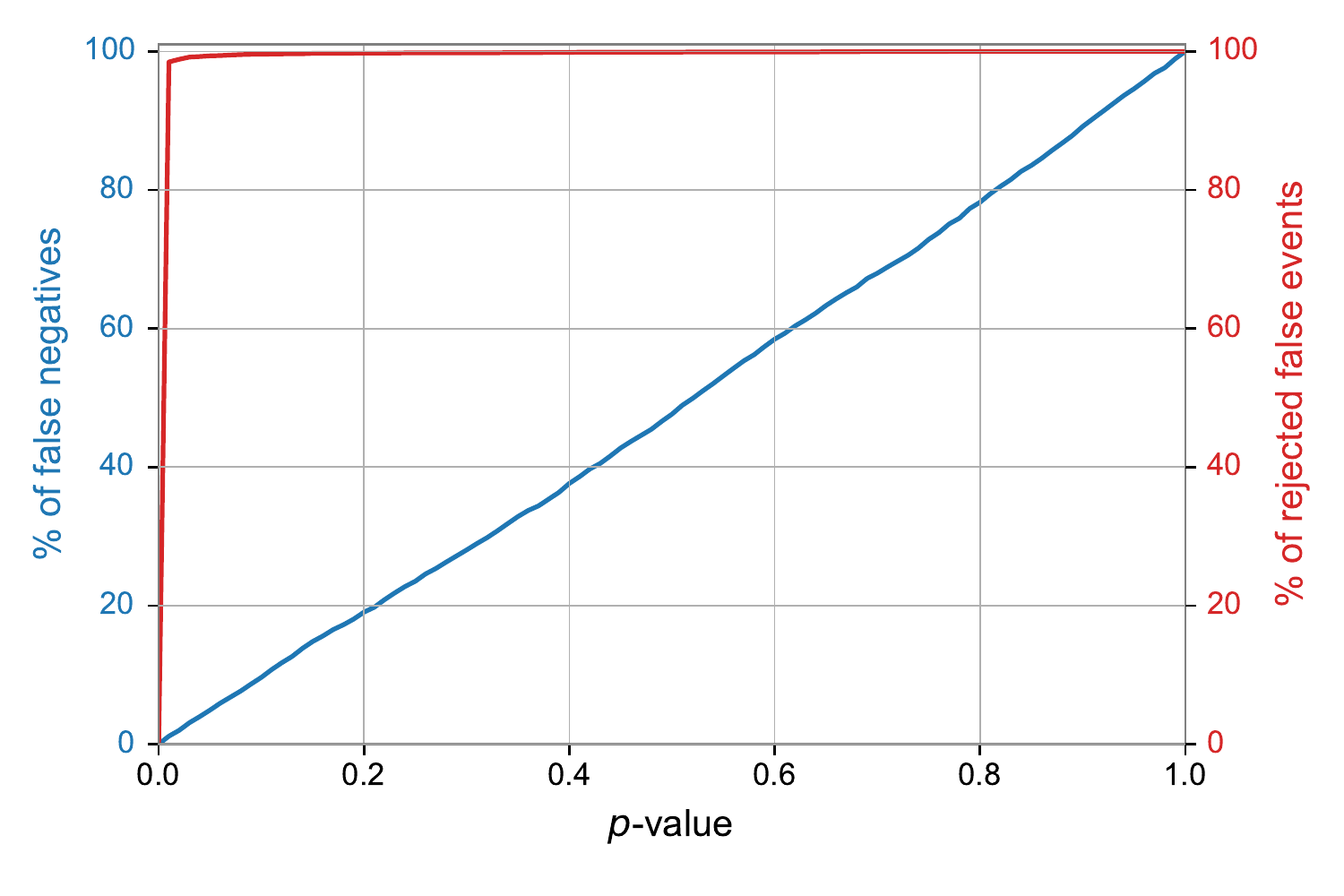}
    \caption{ Blue line: the percent of false-negative signals as a function of $p$-value. This line was determined by simulating domain-wall signals on Gaussian-distributed noise (as per Table.~\ref{table: MagInfo}). Red line: percent of true-negative signals as a function of $p$-value. The true-negative signals were generated similar to the true-positive signals, except with amplitude measurements inconsistent with a domain-wall signal.}
    \label{fig:FNline}
\end{figure}

The $p$-value represents the likelihood that deviations between the amplitudes measured at each sensor and the expected amplitudes corresponding to the most likely domain-wall crossing event (as defined in Sec.~\ref{sec:projection}) can be explained by the characteristic noise of the sensors. A relevant check of the analysis algorithm is the distribution of the false negatives with respect to the $p$-value. Domain-wall signals inserted in Gaussian-distributed noise should exhibit a flat distribution with respect to the $p$-value. This can be seen in blue in Fig.~\ref{fig:FNline}, where the cumulative probability of finding an event is proportional to the $p$-value, confirming the expected behavior.

In contrast, if pulses with random amplitudes are inserted into Gaussian-distributed noise, the $p$-value is generally close to zero; which can be seen in the red line in Fig.~\ref{fig:FNline}. The line is obtained by inserting pulses with timings consistent with a domain-wall crossing but having random amplitudes. This demonstrates how the $p$-value threshold provides a method to distinguish signal patterns matching domain-wall crossing events from spurious non-Gaussian noise (such as random ``spikes'' in the magnetometer data).

\subsection{False-positive analysis}\label{sec:false-positive}

In order to quantify whether a measured signal pattern is sufficiently unlikely to occur due to random noise, it is necessary to study the noise characteristics of the network. The first two stages of the post-selection process are to identify events whose $p$-values are above the designated threshold $p_\text{min}$ and to identify events for which the direction of $\boldsymbol{m}$ matches that of $\boldsymbol{v_\text{scan}}$ within the angular lattice spacing. Inevitably, some events arising from noise may pass the thresholds on $p$-value and directional consistency between $\boldsymbol{m}$ and $\boldsymbol{v}$, so a third threshold characterizing the signal-to-noise of a measurement is introduced. A $5\sigma$ significance for an observed domain-wall crossing event is imposed in order to claim discovery of a domain-wall crossing. This means a probability of about 1~in 1.7~million of being produced by noise over the course of the measurement campaign, $t$.



The number of events above a certain signal-to-noise threshold is expected to follow Poissonian statistics. For a given period of time and number of events detected, a bound with 90\,\% confidence level can be given as rate of false-positives per year. This bound on the false positive rate can be determined by solving Eq.~\eqref{eq:rateBound} for $r_0$, where $n_f$ is determined by simulating $T_\text{samp}$-long data. For events appearing very seldom in the period of time analyzed, the bound is inaccurate because there are not enough events to accurately estimate the underlying rate. This effect is visible when demanding high signal-to-noise ratio events.  However, if one would continue adding data, the rates are expected to continue an exponential trend.

In order to test the exclusion power of the post selection steps, simulated data with Gaussian-distributed noise and random Lorentzian spikes are studied. The data are simulated in 20~min segments. Spikes are inserted randomly with a probability of 10\,\% at each magnetometer with at most one spike per simulated segment. The amplitude takes random values between -20~pT and +20~pT, and the width is fixed to $0.5$~s. The standard deviation of the background noise is extracted from Table~\ref{table: MagInfo}. 

\begin{figure}[ht]
    \centering
    \includegraphics[width=3.2in]{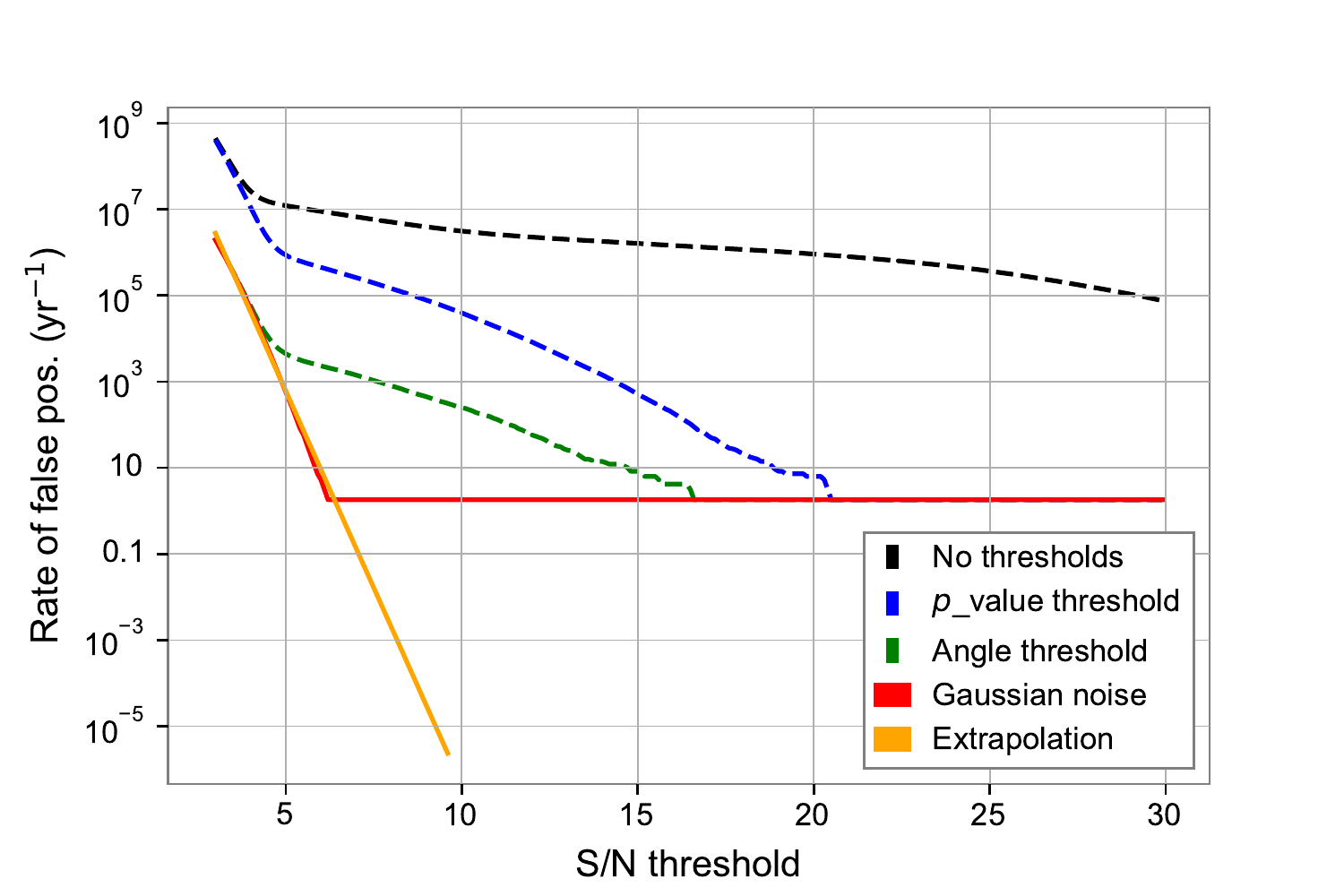}
    \caption{The false-positive analysis for the different stages of the post-selection. The rates are reported as upper-bounds at 90\,\% confidence. The data are composed of Gaussian noise with spurious Lorentzian spikes. The amplitude of the spikes take random values between -20~pT and +20~pT. The black line accounts for all the events, the blue line introduces the $p$-value threshold, the green includes the angle selection. In addition, the orange line shown the extrapolation to $5 \sigma$ significance of detection.}
    \label{fig:FPpostSelection}
\end{figure}

The spikes produce large signal-to-noise events which are shown by the black dotted line in Fig.~\ref{fig:FPpostSelection}. However, because the spikes are unrelated to domain walls, the $p$-value of a spike event is likely to be close to zero. Therefore, a significant amount of high signal-to-noise events can be easily rejected by the $p$-value threshold, as shown by the blue dotted line in Fig.~\ref{fig:FPpostSelection}. The rate of detected events per year is further decreased with the angle threshold as the green dotted line shows. After the post-selection procedure, the rate of false positives is reduced by about four orders of magnitude at a signal-to-noise ratio of~10. For reference, the solid red line indicates the rate of false positives measured with only Gaussian noise background (according to Table~\ref{table: MagInfo}) and no spikes. As expected, it decays exponentially with the signal-to-noise ratio threshold.

For the 1.3~years of simulated data, the most stringent bound on the rate achievable is about 1.8~events per year. However, to determine the threshold for detection, that is, the signal-to-noise ratio resulting in a $5 \sigma$ significance for detection for a measuring time of 1~month, a bound of less than $r_0=6.9\times 10^{-6}$~yr$^{-1}$. is required. To ensure this significance, one would need to create about 4~million times more data than is being analyzed from a measuring campaign. This is computationally impractical, so the false positive rate as a function of the thresholds must be extrapolated to establish the appropriate signal-to-noise ratio threshold. The red solid line in Fig.~\ref{fig:FPpostSelection} is extrapolated with a exponential decay shown by the orange solid line. The $5 \sigma$ significance level is reached for a signal-to-noise ratio of~9.3 when measuring for 1~month. 

A network configuration offers several benefits for detecting domain-wall crossing events and other transient signals associated with beyond-standard-model physics. Since the same event is detected multiple times, a network of sensors offers greater statistical sensitivity compared to only one sensor. Furthermore, the global distribution of the magnetometers along with the GPS-disciplined timing enables accurate characterization of domain-wall-crossing event dynamics. Finally, the combination of the time-domain signal pattern and the pattern of signal amplitudes in the network enables efficient rejection of false-positive events. The rejection of spurious events improve the number of magnetometers taking part in the network.

The identification of plausible events is mainly based on solving Eq.~\eqref{eq:ampLinEq}, a system of linear equations with $n-3=6$ degrees of freedom for $n=9$ magnetometers. When more than four magnetometers are active, the analysis is able to veto events that do not match the expected pattern as described in Sec.~\ref{sec:thresholds}.

\begin{figure}[ht]
    \centering
    \includegraphics[width=6in]{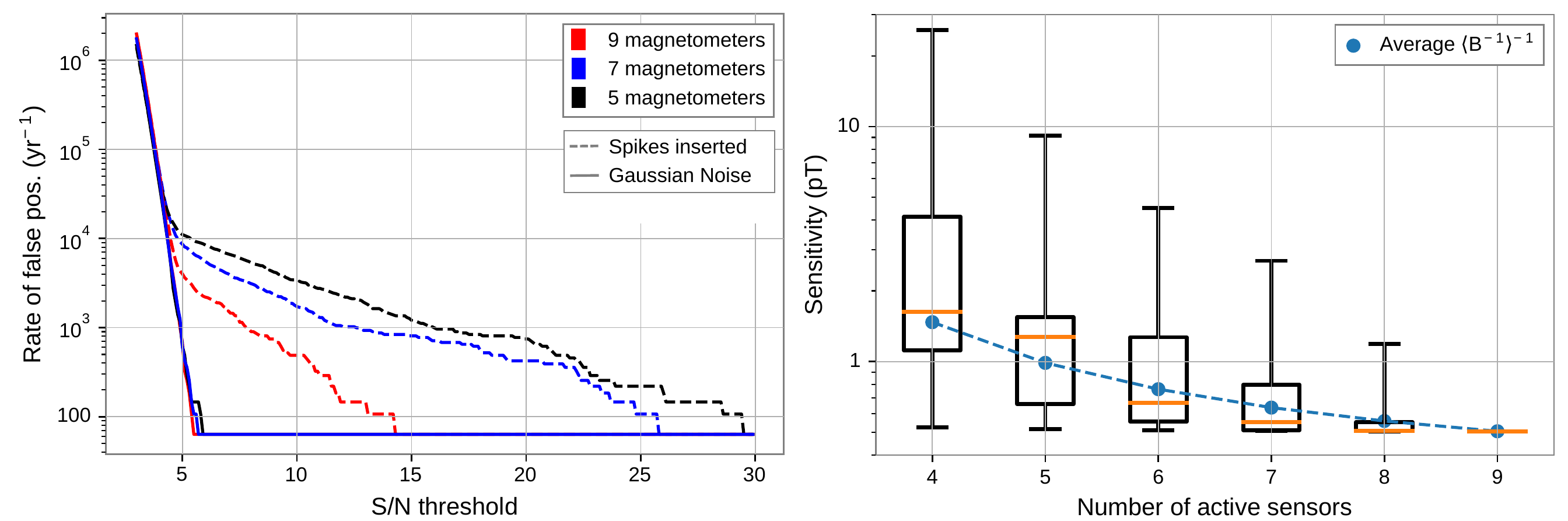}
    \caption{ Left: False-positive analysis with different network sizes. The analysis is performed with Gaussian-distributed noise (solid lines) and Gaussian-distributed noise with spurious spikes injected randomly (dashed lines). Right: sensitivity of the GNOME network when containing different amount of magnetometers active. The box-and-whisker plots are constructed by considering all subsets of the nine magnetometers. The regions of the box-and-whisker indicate 25\,\% of the combinations with the boxes marking the upper- and lower-quartiles separated by the median (orange line).}
    \label{fig:FalsePositiveNmags}
\end{figure}

To test the effects of adding/removing sensors, data were simulated in 1000~samples of 20~min segments. A randomly selected subset of magnetometers is used to simulate the performance of a network with 7 and 5~magnetometers in each sample. Thus the effect of choosing a particular set of magnetometers is averaged out. Apart from the number of magnetometers used, the parameters of the simulation are the same as in Fig.~\ref{fig:FPpostSelection}. The results are shown in Fig.~\ref{fig:FalsePositiveNmags}. The left plot demonstrates the reduction in the rate of false-positive events with additional sensors for background data with random spikes injected (dashed lines). A network of 9~sensors reduces the rate of false-positive events by more than an order of magnitude at a signal-to-noise threshold of~15 as compared to a network of 5~magnetometers. The solid lines show the bound on the false-positives rate for pure, Gaussian-distributed noise. No significant change in the rate of false-positive events on Gaussian-distributed noise is observed with different network sizes because the $p$-value behavior is independent of the number of sensors (or degrees of freedom) in this case. However, there is an improvement on the sensitivity with additional sensors. The right plot in Fig.~\ref{fig:FalsePositiveNmags} shows the $\beta_1$ sensitivity in the least-sensitive direction, as defined in Eq.~\eqref{eq:sensitivity}. Every combination of the nine magnetometers is used to generate the box-and-whisker plots for different sizes of the subsets.

\subsection{Sensitivity}

\begin{figure}[ht]
	\centering
	\includegraphics[width=3.2in]{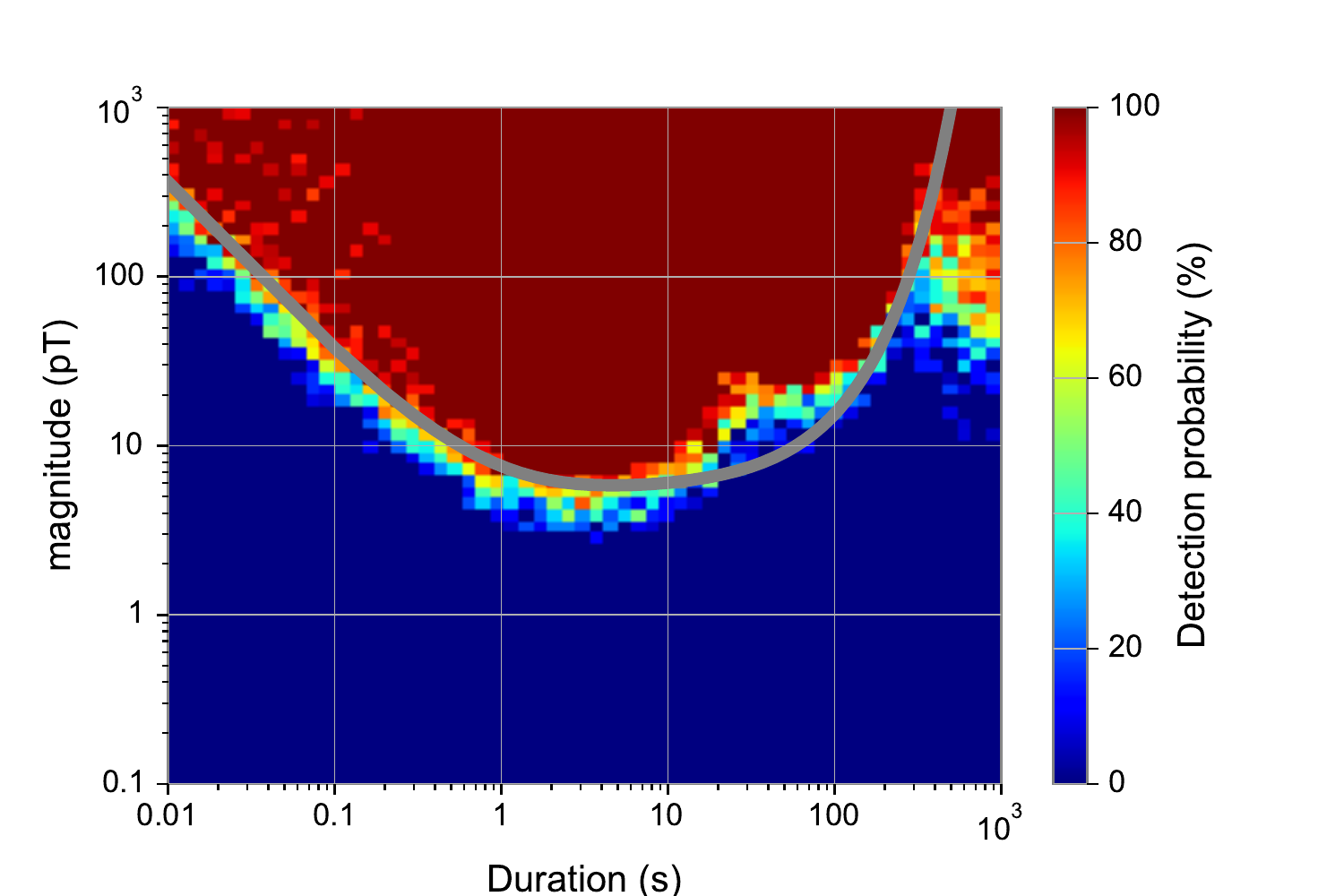}
	\caption{The probability of detection for the algorithm in terms of the  magnitude and duration of a domain-wall crossing signal. This plot is generated with 1~s averaging and a 1/300~Hz high-pass filter. The gray line represents the theoretical limit of detection at 9.3~signal-to-noise ratio.}
	\label{fig:sensitivityPlt}
\end{figure}

The detector network and analysis method determine a class of detectable signals. In particular, the noise of the individual magnetometers, the filters used, and the averaging time determine the duration and magnitude of the detectable signals. The sensitivity is discussed in Sec.~\ref{sec:netSensitivity} with the effects of averaging (binning) and filtering are discussed in \ref{app:filterBin}. For the network characteristics described in Table~\ref{table: MagInfo}, an averaging time of 1~s, 1/300~Hz high-pass filters, and notch filters corresponding to the power line frequencies, the sensitivity to domain-wall signals (assumed to have a Lorentzian shape in the time-domain) is shown by the gray line in Fig.~\ref{fig:sensitivityPlt}.

The ability of the analysis algorithm to observe signals with different durations and magnitudes is studied in 40000~segments of 20~minutes (summarized in Fig.~\ref{fig:sensitivityPlt}). Each of the segments contains a domain-wall signal at random amplitude, duration, direction, and crossing time. The signal is inserted on a Gaussian-distributed noise background defined from the noise characteristics shown in Table~\ref{table: MagInfo}. 

The signal-to-noise limit for accepting the signal is fixed to~9.3, to achieve the false-positive rate needed to reach $5\sigma$ significance for detection of for one month of measurement time (as per Sec.~\ref{sec:false-positive}). If any event is found above the detection limit in the segment, it is marked as a detection. The parameter space is clearly split in two regions: in the lower part the algorithm is unable to identify events, while in the upper part, the events are reliably detected. These two regions are split by the theoretical sensitivity limit. The decrease on the sensitivity for small durations is due to the effects of averaging the data, while the decrease for long signals is due to the high-pass filtering of the data. The deviation from the theoretical line at large signal durations is likely due to poor noise-estimations since the signal spans a time comparable to the segment length. For short signal duration, small errors in aligning the signals through time-shifting leads to significant deviations from the expected amplitude. This results in poor statistical agreement; i.e., a small $p$-value. 

The sensitivity plot is expressed in terms of signal characteristics. However, one is often interested in sensitivity to domain walls in terms of physical parameters. The exact conversion between the signal characteristics and physical parameters depends on the phenomenology being considered. Roughly speaking, the duration of a wall signal is determined by the axion mass (which scales inversely with physical width) and velocity, while the magnitude of the signal is related to the coupling strength. Furthermore, the likelihood that no domain wall was observed because they are rarer than the experiment duration must be considered. These issues will be explored in future publications.

\section{Conclusions}\label{sec:conclusions}
In this work, an analysis algorithm to search for signals in the GNOME data associated with domain-wall crossings was described. The analysis algorithm is designed to look for peaks reproducing the expected timing and amplitude pattern of a domain-wall crossing. The signal pattern is specific to the configuration of GNOME, depending on the geographical location, the alignment of the sensitive axes and the noise characteristics of the magnetometers. The analysis algorithm is demonstrated to effectively discriminate between real domain-wall crossings and noise. The false-positive and false-negative rates for simulated data are analyzed, and a method to evaluate the overall sensitivity of the GNOME network was described. 

The analysis algorithm presented in this work will be applied to the data of the GNOME collaboration. The main challenge with real data is the complexity of the noise characteristics. In order to assess the detection signal-to-noise threshold, the same data being analyzed must be used to estimate the noise. The event rate background will be calculated sampling the data at random times. A real domain-wall signal would not be visible but the noise characteristics would remain. On this incoherent data, the signal-to-noise ratio required for a $5\sigma$~significance detection over the duration of the measurement campaign can be determined. If no events are found above this threshold, the strongest event detection will define the region of exclusion. This will move the gray curve in Fig.~\ref{fig:sensitivityPlt} down and include a larger region of signal characteristics.



\section*{Acknowledgements}
We are sincerely grateful to Chris~Pankow and Josh~Smith for early contributions to the data analysis strategy and Andrei~Derevianko, Ben~Roberts, Conner~Dailey, and Maxim~Pospelov for valuable advice and insights. We are also thankful to all the members of the GNOME collaboration for useful discussions at many stages of development of the analysis procedure, especially H.~Guo, T.~W.~Kornack, W.~Li, S.~Nix, M.~Padniuk, X.~Peng, and D.~Sheng.

This work was supported by the U.S. National Science Foundation under grants PHY-1707875 and PHY-1707803, the Swiss National Science Foundation under grant No. 200021 172686, the European Research Council under the European Union’s Horizon 2020 Research and Innovative Program under Grant agreement No. 695405, the Cluster of Excellence PRISMA+, the National Science Centre, Poland within the OPUS program (Project No. 2015/19/B/ST2/02129), and IBS-R017-D1-2019-a00 of the Republic of Korea.


\bibliographystyle{elsarticle-num}
\bibliography{refs_JAS,references}


\appendix

\section{Filtering/binning effects}\label{app:filterBin}

Filtering and binning will affect both the signal and noise of a signal. The exact nature of these effects will be dependent on the specific characteristics of the signal and noise. Some relevant examples of signals and reasonable approximations of noises will be considered in this appendix. Specialized filters can be used to optimize dark matter searches~\cite{panelli2019applying}, however this appendix will focus on the application of general frequency filters. Further reading related to this appendix can be found in standard textbooks on signal processing, e.g., Ref.~\cite{lathi_signal_1998}.

The effects on noise and signal will be calculated in slightly different ways. In particular, the effects on noise will be calculated with discrete points, while the effects on the signal will be calculated in the continuous limit to simplify the calculation. For these calculations, it helps to define the discrete Fourier transform
\begin{equation}
    (\mathcal{F}_\text{D}f)[k] \equiv \sum_{n=0}^{N-1} f[n] e^{-\frac{2\pi n k}{N} i}
    \quad\text{and}\quad
    (\mathcal{F}_\text{D}^{-1}\tilde{f})[n] = \frac{1}{N}\sum_{k=0}^{N-1} \tilde{f}[k] e^{+\frac{2\pi n k}{N} i}\ ,
\end{equation}
where $f$ is a set of data with $N$ points and $\tilde{f}$ is the Fourier transform. Similarly, the continuous Fourier transform is
\begin{equation}
    \{\mathcal{F}_\text{C} f\}(\omega) \equiv \int_{-\infty}^{\infty}dt\,f(t)e^{-i\omega t}
    \quad\text{and}\quad
    \{\mathcal{F}_\text{C}^{-1} \tilde{f}\}(t) = \frac{1}{2\pi}\int_{-\infty}^{\infty}d\omega\,\tilde{f}(\omega)e^{+i\omega t}\ .
\end{equation}
Note that the discrete Fourier transform is given in terms of frequencies (in units of $r/2N$ where $r$ is the sampling rate) and the continuous transform is given in terms of \emph{angular} frequency, where $\omega\sim 2\pi k$.

\subsection{Effects on noise}
For simplicity, the noise in the sensors will be assumed to be Gaussian distributed. Later, additional approximations will be applied to make the effects easier to calculate. A general frequency filter $\tilde{g}[k]$ will affect the signal according to
\begin{equation}
    f_\text{filt}[n] = \left(\mathcal{F_\text{D}}^{-1}(\tilde{g}\cdot \mathcal{F}_\text{D} f)\right)[n]\ .
\end{equation}
The filter satisfies $\tilde{g}[k]=\tilde{g}^*[N-k]$ so that $f_\text{filt}\in\mathbb{R}$. Note, also, that a circular boundary is assumed for simplicity, so $\tilde{g}[k]=\tilde{g}[N+k]$.

Filtering is a linear operation with a Jacobian matrix given by
\begin{align}
    J_\text{filt}[n,m] \equiv \frac{\partial f_\text{filt}[n]}{\partial f[m]} &= \frac{1}{N} \sum_{k=0}^{N-1} \tilde{g}[k] e^{-\frac{2\pi i}{N}(m-n)} \nonumber \\
    &= \frac{1}{N} \times \begin{cases}
    2\sum_{k=1}^{\frac{N+1}{2}-1} \operatorname{Re}\left( \tilde{g}[k] e^{-\frac{2\pi i}{N}(m-n)} \right) + \tilde{g}[0] & N\ \text{odd}\\
    2\sum_{k=1}^{\frac{N}{2}-1} \operatorname{Re}\left( \tilde{g}[k] e^{-\frac{2\pi i}{N}(m-n)} \right) + \tilde{g}[0] + (-1)^{m-n}\tilde{g}[N/2] & N\ \text{even} 
    \end{cases}\ ,
\end{align}
where $\tilde{g}[k]\in\mathbb{R}$ in the second line which will not shift the signal after filtering. One can show that the Jacobian is a real, circulant (i.e., elements given by the difference between the column and row number), and symmetric matrix. 

For example, consider a simple band-pass filter,
\[
\tilde{g}[k] = \begin{cases}
    1 & k_0 \leq k < k_1\ \text{or}\ N-k_1 < k \leq N-k_0 \\
    0 & \text{else}
\end{cases}\ .
\]
One obtains the Jacobian
\begin{align*}
J_\text{filt}[n,m] = \frac{1}{N}\times & \left[ 
\left\{ \begin{array}{lr}
    \frac{\sin\left( \frac{2\pi}{N} (m-n)(\min\{\lceil \frac{N}{2} \rceil,k_1\}-\frac{1}{2}) \right)-\sin\left( \frac{2\pi}{N} (m-n)(\max\{ k_0, 1 \}-\frac{1}{2}) \right)}{\sin\left( \frac{\pi}{N} (m-n) \right)} & m\neq n \\
    2(\min\{\lceil \frac{N}{2} \rceil,k_1\}-\max\{ k_0, 1 \}) & m=n
\end{array} \right\} \right. \\
&\quad \left. + \tilde{g}[0] + \left\{\begin{array}{lr}
    (-1)^{m-n} \tilde{g}\left[\frac{N}{2}\right] & N\ \text{even} \\
    0 & \text{else}
\end{array}\right\}
\right]\ .
\end{align*}
Likewise, averaging over $T$ points in left-justified bins yields the Jacobian,
\[
J_\text{avg}[n,m] = \begin{cases}
\frac{1}{T} & 0 \leq m-n < T \\
0 & \text{else}
\end{cases}\ .
\]
This is a rolling average, which can be extended by assuming circular boundary conditions on the indices, $n\sim n+N$. The rolling average is equivalent to applying the frequency filter
\[
    \tilde{g}_\text{avg}[k] = \frac{\sin \frac{T\pi}{N}k}{T\sin{\frac{\pi}{N} k}} e^{-i\frac{\pi}{N}k(T-1)}\ .
\]
Observe that the phase is a result of the bins being left-justified and can be removed by using center-justified bins. According to the convolution theorem, the frequency filters can be combined via a product into a single filter. 

If the initial covariance of the data is $\Sigma$, then the covariance in the filtered data is $(\Sigma_\text{filt})[n,m] = \sum_{j,k=0}^{N-1} J_\text{filt}[j,n]\Sigma[j,k]J_\text{filt}[k,m]$. Assuming that the errors are constant and uncorrelated ($\Sigma[m,n] = \sigma^2 \delta_{mn}$), then the resulting covariance is also circulant. This means that the covariance between two points only depends on the distance between those two points. The variance $\bar{\sigma}^2 = \sigma^2 \sum_{j=0}^{N-1} J_\text{filt}[j,0]^2$ is of particular interest. Observe that since $J_\text{filt}$ is symmetric and circulant, $J_\text{filt}[j,0]^2 = J_\text{filt}[j,0]J_\text{filt}[0,j] = J_\text{filt}[j,0]J_\text{filt}[-j,0]$. Thus,
\begin{equation}
    \bar{\sigma}^2 = \frac{\sigma^2}{N} \sum_{k=0}^{N-1} \lvert \tilde{g}[k] \rvert^2\ ,
\end{equation}
equivalent to attenuating the variance by the inner product of the filter, up to a factor of $N$.

\subsection{Effects on the signal}
The effects of the filters on the signal is determined in the continuum limit. The frequency filter $\tilde{g}(\omega)$ on a signal $f(t)$ is given by
\begin{equation}
    f_\text{filt}(t) = \left\{ \mathcal{F}_\text{C}^{-1}\{ \tilde{g}\cdot \mathcal{F}_\text{C} f\} \right\}(t)\ ,
\end{equation}
where $g(-\omega) = g^*(\omega)$ similar to the discrete case. Expanding this equation,
\[
    f_\text{filt}(t) = \frac{1}{\pi} \int_0^\infty d\omega \operatorname{Re}\left( \tilde{g}(\omega) \{\mathcal{F}_\text{C}f\}(\omega) e^{i\omega t} \right)\ ,
\]
where $\{\mathcal{F}_\text{C}f\}(-\omega) = \{\mathcal{F}_\text{C}f\}^*(\omega)$ since $f\in\mathbb{R}$.

For this study, it is useful to consider the case where the signal is Lorentzian,
\begin{equation}
    f(t) = \frac{A}{1 + \left( \frac{t}{\frac{1}{2}\Gamma} \right)^2}\ ,\ \text{so}\quad \{\mathcal{F}_\text{C}f\}(\omega) = \pi A\frac{\Gamma}{2} e^{-\frac{\Gamma}{2} \lvert \omega \rvert}\ .
\end{equation}
Also, a rolling average is accomplished with the frequency filter,
\[
    \tilde{g}_\text{avg}(\omega) = \frac{2}{\omega T}\sin \left(\frac{\omega T}{2}\right)\ .
\]
Likewise, a simple high-pass filter is given by $\tilde{g}_\text{hp}(\omega)=\Theta\left(\lvert\omega\rvert-\omega_\text{L}\right)$.

First, consider applying both a rolling average and a high-pass filter. The resulting signal will be
\[
    f_\text{filt}(t) = \frac{A\Gamma}{2T}\operatorname{Im}\left[ \int_1^\infty \frac{d\nu}{\nu} e^{-\omega_\text{L}\left( \frac{\Gamma}{2}-i\frac{T}{2}-it \right) \nu} + \int_1^\infty \frac{d\nu}{\nu} e^{-\omega_\text{L}\left( \frac{\Gamma}{2}+i\frac{T}{2}-it \right) \nu} \right]\ .
\]
If a frequency filter is applied without averaging,
\[
    f_\text{filt}(t) = \frac{A\Gamma}{2} \frac{\frac{\Gamma}{2}\cos(\omega_\text{L} t) - t\sin(\omega_L t) }{t^2+(\Gamma/2)^2}\ .
\]
Finally, consider averaging without a high-pass filter:
\[
    f_\text{filt}(t) = \frac{A\Gamma}{2T} \left[ \arctan\left( \frac{t+T/2}{\Gamma/2} \right) - \arctan\left( \frac{t-T/2}{\Gamma/2} \right) \right]\ .
\]

When determining by how much the signal is changed, a point is sampled from $f_\text{filt}$. Optimistically, this will be the maximum, $f_\text{filt}(0)$. However, in practice, the data are binned meaning that instead of simply using the rolling average, only one point per bin width is used. This means that in the worst case, the maximum observed value is generally\footnote{Strong filters can cause oscillatory effects in the signal. This means that the largest observed value may not be around the peak, depending on the signal width. For simplicity, signals with widths beyond which this effect occurs can be considered unobservable.} $f_\text{filt}(T/2)$. If, for example, two sets of bins offset by half a bin width is used, this worst-case-scenario improves to $f_\text{filt}(T/4)$.

\end{document}